# Integrated Expansion Planning of Electric Energy Generation, Transmission, and Storage for Handling High Shares of Wind and Solar Power Generation

Mojtaba Moradi-Sepahvand, and Turaj Amraee[*]

*Faculty of Electrical Engineering, K.N. Toosi University of Technology, Tehran, Iran.*

***Abstract***— In this paper, an integrated multi-period model for long term expansion planning of electric energy transmission grid, power generation technologies, and energy storage devices is introduced. The proposed method gives the type, size and location of generation, transmission and storage devices to supply the electric load demand over the planning horizon. The sitting and sizing of Battery Energy Storage (BES) devices as flexible options is addressed to cover the intermittency of Renewable Energy Sources (RESs), mitigate lines congestion, and postpone the need for new transmission lines and power plants installation. For efficient handling of RESs uncertainties, and operational flexibility, the upward and downward Flexible Ramp Spinning Reserve (FRSR) are modeled. Besides, the Low-Carbon Policy (LCP) is considered in the objective function of the proposed Transmission, Generation, and Storage Expansion Planning (TGSEP) model. A hierarchical clustering method that can preserve the chronology of input time series throughout the planning horizon periods is developed to capture the short-term uncertainties of load demand and RESs. The short-term operational flexibility requirements make the joint long-term transmission and generation planning a high computational problem. Therefore, the Mixed-Integer Linear Programming (MILP) formulation of the model is solved using an accelerated Benders Dual Decomposition (BDD) method. The IEEE RTS test system is utilized to validate the effectiveness of the proposed joint expansion planning model.

***Keywords***— *Integrated Electric Energy Generation and Transmission Expansion, Battery Energy Storage Planning, Renewable Power, Low-Carbon Policy, Hierarchical Clustering.*

___________
[*]Corresponding author.
Faculty of Electrical Engineering, K.N. Toosi University of Technology, Tehran, Iran, Postal Code:1631714191.
E-mail: amraee@kntu.ac.ir

## NOMENCLATURE

*Indices:*

| | |
|---|---|
| $s, h$ | Index of planning stages, index of representative hours. |
| $j, g$ | Index of all buses, index of all types of new candidate and existing Thermal Generation Units (TGUs). |
| $l, c$ | Index of all lines, index of the allowable candidate lines in a corridor. |
| $t, p$ | Index of all types of BES devices, index of linear segments of the TGUs operation cost function. |
| $d, v$ | Index of the allowable candidate TGUs in a bus, index of the TGU technology types. |

*Sets:*

| | |
|---|---|
| $\Omega_S, \Omega_H$ | Sets of planning stages, and representative hours. |
| $\Omega_B, \Omega_G$ | Sets of all buses, and generators. |
| $\Omega_{BS}, \Omega_W, \Omega_{PV}$ | Sets of candidate buses for installing Battery Energy Storage (BES) devices, wind and Photovoltaic (PV) farms. |
| $\Omega_{NU}, \Omega_{EU}, \Omega_U$ | Sets of all types of new candidate and existing TGUs, as subsets of $\Omega_U$ (set of all units). |
| $\Omega_{el}, \Omega_{nl}, \Omega_{nc}$ | Sets of all existing lines, candidate lines, and candidate lines in new corridors as a subset of $\Omega_{nl}$. |
| $\Omega_C, \Omega_T, \Omega_P$ | Sets of the allowable candidate lines in a corridor, all types of BES devices, and linear segments of the TGUs operation cost function. |
| $\Omega_D, \Omega_{TT}$ | Sets of the allowable candidate TGUs in a bus, and the TGU technology types. |

*Parameters:*

| | |
|---|---|
| $S, H, P$ | Total number of planning stages, representative hours, and linear segments of the TGUs operation cost function. |
| $i, T$ | Interest rate and Lifetime of equipment (year). |
| $CL_l$ | Investment cost of new line $l$ including the cost of conductors of single/double-circuits and towers ($10^6$ \$/km). |
| $CT_{j,g}$ | Investment cost of new candidate TGUs ($10^6$ \$/MW). |
| $CW_j, CV_j$ | Investment cost of new wind and PV farms ($10^6$ \$/MW). |
| $Cap_{j,g}$ | Power capacity of each thermal generation unit $g$, in bus $j$ (MW). |
| $Rt_l$ | Right of Way (RoW) cost of line $l$ including land cost ($10^6$ \$/Km). |
| $L_l, SS_l$ | Line length of all new lines (Km), and new substation cost in new corridors ($10^6$ \$). |
| $FP_{j,g}^p, HR_{j,g}^p$ | Fuel Price for generated power in segment $p$ of each thermal generation unit $g$, in bus $j$ (\$/MBTU), Heat Rate of thermal generation unit $g$, in bus $j$, according to the generated power in each segment $p$ (MBTU/MWh). |
| $NE_{j,g}, MC_{j,g}$ | Number of existing thermal generation units $g$, in bus $j$, and Maintenance cost of TGU $g$, in bus $j$ (\$/MW-year). |
| $MCW_j, MCP_j$ | Maintenance cost of wind and PV farms in bus $j$ ($/MW-year). |
| $EM_{j,g}, EC_{j,g}^p$ | CO2 emission of each thermal generation unit $g$, in bus $j$ (ton/MWh), CO2 emission cost for generated power in segment $p$ of each thermal generation unit $g$, in bus $j$ ($/ton). |
| $TN_{j,g}$ | Maximum allowable number of new thermal generation unit $g$, in bus $j$, and at each stage, as tunnel limit. |
| $CLS_{j,h}, CWC_{j,h}$ | Load shedding and wind curtailment penalty costs ($/MWh) in bus $j$, and hour $h$. |
| $RPU_{j,g}, RPD_{j,g}$ | Ramp up and ramp down limits of thermal generation unit $g$, in bus $j$ (MW/hr). |
| $RUW_{j,g}, RDW_{j,g}$ | Maximum upward and downward Flexible Ramp Spinning Reserve (FRSR) capacity of thermal generation unit $g$, in bus $j$ (MW). |
| $TRR^{min}$ | Total minimum required reserve at each planning stage as a certain percentage of the system peak load. |
| $CX_{s,v}^{min}, CX_{s,v}^{max}$ | Minimum and maximum expected contribution of the technology type $v$ in total available capacity at stage $s$. |
| $MX_{j,g,v}$ | Discriminant matrix of the technology type $v$ of thermal generation unit $g$, in bus $j$. |
| $\alpha, \beta$ | Expected contribution of RESs in supplying the total energy demand at the end of planning horizon based on (Renewable Portfolio Standard) RPS policy, and maximum allowable total wind curtailment at each stage of the planning horizon. |
| $\gamma, \Phi$ | Maximum permitted load shedding in each bus, and maximum permitted total load shedding in each stage of the planning horizon. |
| $\{\bullet\}^{max}, \{\bullet\}^{min}$ | Maximum/minimum limits of bounded variables. |
| $A, K$ | Directional Connectivity matrices of existing and new constructed lines with buses. |
| $SC_{j,t}, CC_{j,t}, Cd_{j,t}$ | Investment cost of energy capacity ($/MWh) and power capacity ($/MW), and degradation cost ($/MWh) for each BES of type $t$ in bus $j$. |
| $SM_{j,t}, CM_{j,t}$ | Maximum energy (MWh) and power (MW) capacity of BES of type $t$ in bus $j$. |
| $\eta_c, \eta_d$ | Charging and discharging efficiency of BES devices. |
| $Wf_h, PVf_h, Lf_h$ | Hourly representative factors of wind and PV farms output power, and load demand. |
| $LD_j^{PK}, LG_s$ | Peak load of bus $j$ (MW), Load growth factor at stage $s$. |
| $\rho_h$ | The weight of extracted representative hour $h$. |
| $RC_{j,g}$ | The upward and downward FRSR cost ($/MWh). |
| $\mathcal{M}, \Psi, B_l$ | Big-M, Base power of the system (MVA), Per unit susceptance of all lines. |

***Binary Variables:***

| | |
|---|---|
| $X_{s,j,g,d}, Y_{s,l,c}$ | Binary variable for construction of $d^{th}$ candidate thermal generation unit $g$, in bus $j$, at stage $s$, Binary variable for construction of candidate line $l$ at stage $s$, in corridor $c$ (equals 1 if the candidate line is constructed and 0 otherwise). |
| $I_{s,j,t}, U_{s,j,t,h}$ | Binary variables of installing BES of type $t$, in bus $j$, at stage $s$, and charging or discharging state of BES of type $t$, in bus $j$, at stage $s$, and hour $h$. |

*Positive Continuous Variables:*

| | |
|---|---|
| Z | Total Planning Cost ($TC_P$). |
| $N_{s,j,g}$ | Total number of constructed thermal generation unit $g$ in bus $j$ at stage $s$. |
| $P_{s,j,g,h}, PS_{s,j,g,h,p}$ | Power output of thermal generation unit $g$, in bus $j$, at stage $s$, and hour $h$, Power generation of segment $p$ of thermal generation unit $g$, in bus $j$, at stage $s$, and hour $h$ (MW). |
| $Pw_{s,j}, PC_{s,j,h}$ | Total power capacity of installed wind farm in bus $j$, and at stage $s$ (MW), and wind curtailment of installed wind farm in bus $j$, at stage $s$, and hour $j$ (MW). |
| $Pv_{s,j}$ | Total power capacity of PV farm $j$ at stage $s$ (MW). |
| $RU_{s,j,g,h}, RD_{s,j,g,h}$ | Upward and Downward FRSR of thermal generation unit $g$, in bus $j$, at stage $s$, and hour $h$ (MW). |
| $LS_{s,j,h}$ | Load shedding in bus $j$ at stage $s$ and hour $h$ (MW). |
| $Pd_{s,j,t,h}, Pc_{s,j,t,h}$ | Discharging and charging power of BES of type $t$, in bus $j$, at stage $s$, and hour $h$ (MW). |
| $E_{s,j,t,h}$ | Stored energy of BES of type $t$, in bus $j$, at stage $s$, and hour $h$ (MWh). |

*Free Continuous Variables:*

| | |
|---|---|
| $Pe_{s,l,h}, Pl_{s,l,c,h}$ | Power flow of existing line $l$, at stage $s$, and hour $h$ (MW), Power flow of new constructed line $l$, in corridor $c$, at stage $s$, and hour $h$ (MW). |
| $\theta_{s,j,h}$ | Voltage angle of bus $j$, at stage $s$, and hour $h$. |

*Compact Representation:*

| | |
|---|---|
| **Y** | Vector of binary decision variables. |
| **R** | Vector of wind and PV farms power capacity variables (as positive continuous variables). |
| **P** | Vector of positive continuous operational variables. |
| **F** | Vector of free continuous variables. |

## I. INTRODUCTION

### 1. BACKGROUND AND LITERATURE REVIEW

Due to the rapid electric load demand growth and economic or environmental restrictions, the power system expansion should be planned using modern tools such as Renewable Energy Sources (RESs) and Battery Energy Storage (BES) devices. The existing transmission lines more often are not able to transfer the required power to the demand side. In this regard, the Transmission Expansion Planning (TEP) and other supporting tools, e.g., FACTS and BES devices, are taken into account. Growing load demand, and considering Low Carbon (LC) regulations result in national and global commitments to Renewable Portfolio Standard (RPS) Policies to decrease the carbon emission from fossil fuels combustion. To realize RPS and LC policies, the long-term Generation Expansion Planning (GEP) studies should be reshaped. The growing penetration of RESs under the RPS policy, which are often placed in remote areas, needs new transmission line construction. Therefore, the joint planning of GEP and TEP problems is crucial because the sitting and sizing of new generation units is affected by the transmission expansion plans [1]. Due to RES generation intermittency and coordination between generation and transmission systems, the GEP and TEP problems should be integrated to optimize energy utilization [2].

Conventionally GEP and TEP studies are carried out separately. In [1] and [3], detailed reviews of the recent GEP problem models are presented. A multi-stage model for the GEP problem is proposed in [4] to investigate the transition toward high penetration of RESs. The main concern in [4] is capturing the operational challenges of the increased share of RESs in the power system. In [5], a bilevel GEP model is developed to merge the RESs market into the power system operation according to the RPS policy. A multi-stage and multi-objective GEP model is presented in [6] considering carbon emission and operational flexibility.

A comprehensive review of recent TEP models is provided in [7] and [8]. In [9], a reduction method for search space of TEP problem based on DC Optimal Power Flow (DC-OPF) is presented in which the main purpose is limiting the number of candidate lines. In [10], a DC-OPF based model considering AC transmission lines conversion to DC lines for TEP problem is presented. The high penetration of RES along with technical and economic influences of ES devices are considered in the proposed multi-year model in [10]. A single-year robust model based on DC-OPF for the TEP problem is proposed in

[11], considering uncertainty sets of renewable output. A risk-based single-year model is presented in [12] to minimize the wind power curtailment in the TEP problem. In [13], a hurricane resilient TEP model considering the siting and sizing of BES and wind farms is proposed. Due to the high cost of constructing new right of way, and environmental issues, upgrading the capacity of existing transmission lines is analyzed in [10, 14-16]. In order to increase the existing transmission line capacity by 10% to 30%, the Dynamic Thermal Rating (DTR) system is considered in [14-16]. In [14], the simultaneous deployment of BES devices, demand response, and DTR system for peak load supplying and enhancing the reliability of transmission network is addressed. Also, the influence of DTR system and demand response on minimizing the size of required BES is investigated. To improve the operation of transmission lines equipped with DTR system, a risk-based management framework is proposed in [15]. In order to investigate both the aleatory and epistemic uncertainties of end-of-life failure model of transmission lines equipped with the DTR system, a hybrid method is developed in [16], that can improve the system reliability evaluation. A review of DTR impacts on improving the power system reliability is presented in [17].

Although the previous research works (i.e., [4-6, 9-13]) present several interesting models for GEP and TEP in a separate way, some recent investigations, e.g., [2, 18-25], work on integrating GEP and TEP as interrelated problems to present more beneficial models. The available integrated GEP and TEP models can be analyzed regarding some unique characteristics such as the high penetration of RES, the procedure of uncertainty handling, utilizing ES devices, and the planning time horizon (i.e., single or multi-period). An integrated multi-year GEP and TEP model considering wind power and demand response impacts is proposed in [2]. Although the ES devices and short-term and long-term uncertainties are ignored, the tractability of the model complexity is a challenging issue in the case of large-scale systems. A study on the development of North Sea region power system architecture, including onshore and offshore generating units, and offshore transmission lines to integrate high shares of RESs, is conducted in [18]. This study is performed for a multi-period planning horizon without considering the power system uncertainties and ES devices. In [19], a joint multi-year GEP and TEP model is proposed considering renewable based distributed generation and $CO_2$ emission. Only demand uncertainty is considered in [19], ignoring ES devices. In [20], the power system long-term and short-term uncertainties, and RES penetration are considered in a stochastic adaptive robust model for GEP and TEP problems. The proposed model is solved for a single target year without modeling ES devices. A coordinated multi-year GEP and TEP problem considering the hourly net load ramping and annual net load duration curve uncertainties is modeled in [21]. In the presented model, FACTS devices are utilized to delay the construction of generators and transmission lines under wind penetration without considering the impacts of ES devices. In [22], a static co-planning of GEP and TEP under wind power integration is investigated by utilizing Weibull and Normal distribution to deal with wind power and load demand uncertainties, respectively. In the presented model, the ES devices are not considered. In addition to wind farms, concentrated solar Photovoltaic (PV) units are considered in a stochastic multi-year multi-objective coordinated GEP and TEP model in [23], and the load and RESs uncertainties are handled using a scenario based approach without considering the impacts of ES devices. In [24], GEP and TEP problems are co-optimized considering the siting and sizing of utility-scale PV farms. In the presented multi-year model, the impacts of ES devices are not considered. An integrated multi-year GEP and TEP model in interconnected power systems considering renewable penetration, and $CO_2$ emission is presented in [25]. Both uncertainty handling and ES devices are ignored in [25].

By increasing RESs penetration in the power system with intermittent output power, some technical challenges in the power system are appeared. In 41 states of the USA, about 82 GW wind farms are constructed [26], and in China, it is anticipated that the installed wind farms capacity reaches 400 GW by 2030. To deal with the uncertainties of RESs power and load demand, considering ES devices in power system planning is essential. Based on recent developments in ES technologies, the GEP and TEP models can be more optimized. By considering ES devices, the need for new thermal generation units (TGUs), and consequently $CO_2$ emission, the transmission lines congestion, and the RESs power curtailment can be minimized [27]. The applications of ES devices in the power system are growing recently, e.g., utilizing ES is expected to exceed 35 GW in the USA by 2025 [27]. Also, according to International Energy Agency (IEA), to reduce global warming by two degrees of Celsius, the install capacity of ES is needed to reach 450 GW in 2050 [28]. In [28] a comprehensive review on the recent ES utilization trends in the power system is discussed. In several research works, the impacts of ES devices on the separate GEP (e.g., [4, 29, 30]), and TEP (e.g., [10, 13, 26, 27, 31]) problems are investigated.

In [4] and [29], the GEP problem considering ES devices under RESs penetration is presented. In [29], a chronological time-period clustering method is introduced to capture load and wind power uncertainties. An expansion planning model for generation (including TGU and RES) and ES devices based on a power-based UC formulation is presented in [30]. The role of ES technologies in renewable energy portfolio planning is investigated in [32]. In [26] and [27], the co-planning of transmission lines and ES devices under wind farms penetration is addressed. In [26], the proposed model considers both long and short-term uncertainties, while a security constrained model concerning short-term uncertainty is presented in [27]. In [31], coordinated planning of transmission, ES, and wind farms under RPS policy, considering line switching is investigated. The sitting and sizing of BES devices to reduce the RESs curtailment considering power flow constraints is proposed in [33].

Under the integration of ES devices and penetration of RESs, an important issue in the long-term planning models is capturing the short-term uncertainties of load demand and RES output power, to handle the power system operation more accurately. Due to the high computational burden and complexity, it is impossible to consider each year whole hours in the planning horizon. Therefore, it is required to develop some methods for dealing with the short-term variation of load demand, and intermittency of RES output power with an acceptable computational burden. One of the primary methods is using the Load Duration Curve (LDC) and approximate it with some blocks to capture the uncertainty of load. Although the complexity of problem is decreased, the accuracy of this method is low. Most of the investigations are based on the stochastic programming framework, which is a scenario-based method (e.g., [13, 20, 23, 26, 27, 29-31]). In this method, by utilizing a large number of scenarios, the uncertainty is captured, while the complexity by considering a large number of scenarios is remained. Therefore, it is necessary to extract several effective representative scenarios. The common clustering algorithms, like k-means (e.g., [20, 23, 26, 27, 31]), k-medoids (e.g., [30]), and hierarchical clustering (e.g., [13, 29]), are accurate methods that can be utilized to extract proper representative intervals. Contrary to considering the approximated LDC, in clustering algorithms the parameters chronology will be considered as much as possible.

## 2. RESEARCH GAPS AND CONTRIBUTIONS

In most previous research works, the operational flexibility requirements in joint GEP and TEP studies are ignored. Additionally, in existing integrated GEP and TEP models (e.g., [2, 18-25]), the lack of modern tools such as BES devices for obtaining an optimal planning scheme is still a noticeable gap. Besides, in some previous works, the planning is modeled for just one given target year, so the construction time of new devices, and proper calculation of operational cost are ignored. Regarding these gaps, the main contributions of this work are summarized as follows.

1) In this paper, an integrated co-planning model is developed for multi-stage expansion of transmission network, generation technologies, and BES devices named TGSEP model. The high penetration of renewable resources (including wind and PV farms) is assumed according to the RPS policy. A hierarchical clustering method is developed to capture the uncertainty and variability of load demand and renewable generation with preserving the chronology of input time series throughout all stages of the planning horizon.

2) The upward and downward Flexible Ramp Spinning Reserve (FRSR) is incorporated in the proposed TGSEP model to provide operational flexibility. The sitting and sizing of BES devices as flexible options to maximize the RESs integration, cover the intermittency of RESs, mitigate lines congestion, and postpone the transmission network and generation expansion plans is also incorporated. The Low-Carbon Policy (LCP) for decreasing $CO_2$ emission is considered in the proposed TGSEP model objective function.

3) In order to solve the Mix-Integer Linear Programming (MILP) formulation of the proposed TGSEP model with a tractable complexity, an accelerated Benders Dual Decomposition (BDD) method is utilized.

**3. PAPER STRUCTURE**

The remaining of this paper is organized as follows. Section II describes the details of the proposed integrated co-planning formulation. The proposed TGSEP model overall structure and the developed hierarchical clustering algorithm are presented in Sections III and IV. Section V presents the obtained numerical results, and finally, the conclusions are provided in section VI.

## II. THE PROPOSED FORMULATION

This section presents the complete formulation of the proposed multi-stage TGSEP model in general and BDD framework.

### 1. General Framework

The objective function is presented in (1), in which the Discounted Present Values (DPV) of investment, operation, maintenance, and CO2 emission costs are minimized. The DPV of operation, maintenance, and CO2 emission costs are assumed at the end of each planning stage (each stage contains two years), while the investment costs are considered at the beginning of each planning stage. In (1), as illustrated in Fig. 1, the compact form of objective function including Total Investment Cost ($TC_{Inv}$), Total Operation Cost ($TC_O$), Total Maintenance Cost ($TC_M$), and Total CO2 Emission Cost ($TC_E$) is presented. According to (2), the $TC_{Inv}$ contains the DPV investment cost of all new constructed transmission lines (i.e., ICL), TGUs (i.e., ICT), RESs (i.e., ICR), and BES devices (i.e., ICS), as shown in Fig. 1. All of the DPV investment costs are converted to Equivalent Annual Cost (EAC) using Capital Recovery Factor (CRF) [34]. The EAC for investment cost of new lines, including conductors, towers, and the RoW cost, is represented by ICL in (3). In new corridors, the substation cost is considered just for the first constructed corridor. In (4), the EAC for investment cost of new TGUs is formulated as ICT based on new installed generation capacity. According to (5) and (6), the EAC for investment costs of RESs, including wind and PV farms, and BES devices are formulated by ICR and ICS, respectively.

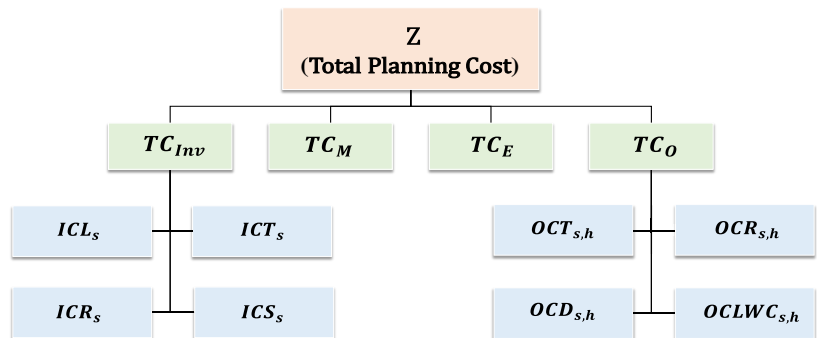

Fig. 1. The structure of proposed model objective function

The DPV of $TC_O$ contains the linearized cost function of TGUs (i.e., OCT) based on the product of heat rate and fuel price, cost of required upward and downward FRSR (i.e., OCR), degradation cost of BES (i.e., OCD) as a result of the charging and discharging cycles, and load shedding (or Energy Not Served (ENS)) and wind curtailment costs (i.e., OCLWC), as expressed in (7) and Fig. 1. In (8)-(11), the OCT, OCR, OCD, and OCLWC are presented. The DPV of $TC_M$ as the summation of the maintenance cost of TGUs (i.e., MCT), and the maintenance cost of wind and PV farms (i.e., MCR) is formulated in (12). The MCT and MCR are expressed in (13) and (14). The MCT contains the maintenance cost of existing and new constructed TGUs. In (15), the DPV of TGUs $TC_E$ is formulated.

$$Min \ Z = TC_{Inv} + TC_O + TC_M + TC_E \tag{1}$$

$$TC_{Inv} = \sum_{s \in \Omega_S} \left[ \left( \frac{2}{(1+i)^{2s-1}} \right) \times [ICL_s + ICT_s + ICR_s + ICS_s] \right] \tag{2}$$

$$ICL_s = \frac{10^6 \times i(1+i)^{T_{Line}}}{(1+i)^{T_{Line}}-1} \times \left[ \sum_{l \in \Omega_{nl}} (IC_l + Rt_l).L_l \times (\sum_{c \in \Omega_C} Y_{s,l,c}) + \sum_{l \in \Omega_{nc}} [Y_{s,l,c=1}.(SS_l)] \right] \tag{3}$$

$$ICT_s = \sum_{j \in \Omega_G} \sum_{g \in \Omega_{NU}} \frac{10^6 \times i(1+i)^{T_{Gen.g}}}{(1+i)^{T_{Gen.g}}-1} \times [(CT_{j,g}.Cap_{j,g}) \times N_{s,j,g}] \tag{4}$$

$$ICR_s = \frac{10^6 \times i(1+i)^{T_{wind}}}{(1+i)^{T_{wind}}-1} \times \left[ \sum_{j \in \Omega_W} CW_j \ .Pw_{s,j} \right)] + \frac{10^6 \times i(1+i)^{T_{PV}}}{(1+i)^{T_{PV}}-1} \times \left[ \sum_{j \in \Omega_{PV}} CV_j \ .Pv_{s,j} \right)] \tag{5}$$

$$ICS_s = \sum_{j \in \Omega_{BS}} \sum_{t \in \Omega_T} \frac{i(1+i)^{T_{ES_t}}}{(1+i)^{T_{ES_t}}-1} \times (I_{s,j,t}) \times [(SC_{j,t}.SM_{j,t}) + (CC_{j,t}.CM_{j,t})] \tag{6}$$

$$TC_O = \sum_{s \in \Omega_S} \left[ \left( \frac{2}{(1+i)^{2s}} \right) \times \sum_{h \in \Omega_H} 8760.\rho_h \times [OCT_{s,h} + OCR_{s,h} + OCD_{s,h} + OCLWC_{s,h}] \right] \tag{7}$$

$$OCT_{s,h} = \sum_{j \in \Omega_G} \sum_{g \in \Omega_U} (\sum_{p \in \Omega_P} [(FP_{j,g}^p . HR_{j,g}^p . PS_{s,j,g,h,p})]) \tag{8}$$

$$OCR_{s,h} = \sum_{j \in \Omega_G} \sum_{g \in \Omega_U} (RC_{j,g} \times [RU_{s,j,g,h} + RD_{s,j,g,h}]) \tag{9}$$

$$OCD_{s,h} = \sum_{j \in \Omega_{BS}} \sum_{t \in \Omega_T} (Cd_{j,t}.Pd_{s,j,t,h}) \tag{10}$$

$$OCLWC_{s,h} = \sum_{j\in\Omega_B}(CLS_{j,h}.LS_{s,j,h}) + \sum_{j\in\Omega_W}(CWC_{j,h}.PC_{s,j,h}) \tag{11}$$

$$TC_M = \sum_{s\in\Omega_S}\left[\left(\frac{2}{(1+i)^{2s}}\right) \times [MCT_s + MCR_s]\right] \tag{12}$$

$$MCT_s = \sum_{j\in\Omega_G}\left[\sum_{g\in\Omega_{EU}}([MC_{j,g} \times Cap_{j,g}].NE_{j,g}) + \sum_{g\in\Omega_{NU}}([MC_{j,g} \times Cap_{j,g}].N_{s,j,g})\right] \tag{13}$$

$$MCR_s = \sum_{j\in\Omega_W}(MCW_j.Pw_{s,j}) + \sum_{j\in\Omega_{PV}}(MCP_j.Pv_{s,j}) \tag{14}$$

$$TC_E = \sum_{s\in\Omega_S}\left[\left(\frac{2}{(1+i)^{2s}}\right) \times \sum_{h\in\Omega_H} 8760.\rho_h \times \left[\sum_{j\in\Omega_G}\sum_{g\in\Omega_U} EM_{j,g} \times \left(\sum_{p\in\Omega_P}[EC_{j,g}^p.PS_{s,j,g,h,p}]\right)\right]\right] \tag{15}$$

The constraints of (16) to (23) are introduced for modeling the TGUs operational limits. In (16), the bounds of linear power generation segments are formulated as the division of available (i.e., the summation of existing and new constructed) capacity of each unit by the total number of segments. In (17), the hourly power generation of each TGU is considered as a set of linear power generation segments to linearize the nonlinear generation cost function. In (18), the maximum power generation of each TGU considering the upward FRSR is enforced. The upper bound of downward FRSR is formulated in (19). In (20), the total number of new constructed TGUs is introduced as the summation of all correspond construction binary variables. The constraint of (21) guarantees that the constructed TGU at each stage will remain until the end of planning horizon. Due to restrictions in constructing TGUs at each stage of the planning horizon, the tunnel limit constraint is introduced as given in (22). With regard to the upward and downward FRSR, in each hour, the ramping constraints of TGUs are presented in (23).

$$0 \leq PS_{s,j,g,h,p} \leq \left([NE_{j,g} + N_{s,j,g}] \times Cap_{j,g}\right)/P \quad \forall s\in\Omega_S, j\in\Omega_G, g\in\Omega_U, h\in\Omega_H, p\in\Omega_P \tag{16}$$

$$P_{s,j,g,h} = \sum_{p\in\Omega_P} PS_{s,j,g,h,p} \quad \forall s\in\Omega_S, j\in\Omega_G, g\in\Omega_U, h\in\Omega_H \tag{17}$$

$$P_{s,j,g,h} + RU_{s,j,g,h} \leq [NE_{j,g} + N_{s,j,g}] \times Cap_{j,g} \quad \forall s\in\Omega_S, j\in\Omega_G, g\in\Omega_U, h\in\Omega_H \tag{18}$$

$$RD_{s,j,g,h} \leq P_{s,j,g,h} \quad \forall s\in\Omega_S, j\in\Omega_G, g\in\Omega_U, h\in\Omega_H \tag{19}$$

$$N_{s,j,g} = \sum_{d\in\Omega_D} X_{s,j,g,d} \quad \forall s\in\Omega_S, j\in\Omega_G, g\in\Omega_{NU} \tag{20}$$

$$X_{s-1,j,g,d} \leq X_{s,j,g,d} \quad \forall s\in\Omega_S, j\in\Omega_G, g\in\Omega_{NU}, d\in\Omega_D \tag{21}$$

$$N_{s,j,g} - N_{s-1,j,g} \leq TN_{j,g} \qquad \forall s \in \Omega_S, j \in \Omega_G, g \in \Omega_{NU} \tag{22}$$

$$\begin{cases} P_{s,j,g,h} + RU_{s,j,g,h} - P_{s,j,g,h-1} \leq RPU_{j,g} & \forall s \in \Omega_S, j \in \Omega_G, g \in \Omega_U, h \in \Omega_H \\ P_{s,j,g,h-1} + RD_{s,j,g,h} - P_{s,j,g,h} \leq RPD_{j,g} & \forall s \in \Omega_S, j \in \Omega_G, g \in \Omega_U, h \in \Omega_H \end{cases} \tag{23}$$

The installed capacity of all TGUs, wind, and PV farms must be adequate to supply the hourly load demand at each planning horizon stage. In this regard, the total available power generation capacity at each planning horizon stage is considered to be equal or greater than the summation of system peak load and a minimum required reserve (as a certain percentage of system peak load), as presented in (24).

$$(1 + TRR^{min}) \times \left( (1 + LG_s)^s \cdot \sum_{j \in \Omega_B} LD_j^{pk} \right) \leq \sum_{j \in \Omega_G} \sum_{g \in \Omega_U} \left( [NE_{j,g} + N_{s,j,g}] \times Cap_{j,g} \right)$$
$$+ \sum_{j \in \Omega_W} Pw_{s,j} + \sum_{j \in \Omega_{PV}} Pv_{s,j} \qquad \forall s \in \Omega_S \tag{24}$$

Due to the utilization of different thermal generation technologies, the capacity mix constraints are defined to adjust the maximum and minimum available capacity for each type of TGUs technology at each stage of planning horizon, as presented in (25) and (26).

$$\sum_{j \in \Omega_G} \sum_{g \in \Omega_U} MX_{j,g,v} \times \left( [NE_{j,g} + N_{s,j,g}] \times Cap_{j,g} \right) \leq CX_{s,v}^{max} \times \left[ \sum_{j \in \Omega_G} \sum_{g \in \Omega_U} \left( [NE_{j,g} + N_{s,j,g}] \times Cap_{j,g} \right) \right] \qquad \forall s \in \Omega_S, v \in \Omega_{TT} \tag{25}$$

$$CX_{s,v}^{min} \times \left[ \sum_{j \in \Omega_G} \sum_{g \in \Omega_U} \left( [NE_{j,g} + N_{s,j,g}] \times Cap_{j,g} \right) \right] \leq \sum_{j \in \Omega_G} \sum_{g \in \Omega_U} MX_{j,g,v} \times \left( [NE_{j,g} + N_{s,j,g}] \times Cap_{j,g} \right) \qquad \forall s \in \Omega_S, v \in \Omega_{TT} \tag{26}$$

The upward and downward FRSR are modeled to deal with the uncertainty of load demand and RESs output power, and provide flexibility requirements, as given in (27) to (30). In (27) and (28), the hourly upper bound of upward and downward FRSR are defined, respectively. As given in (29) and (30), the lower bound for the total hourly upward and downward FRSR are assumed to be 5% and 3% of the total expected RESs output power and the system load, respectively [10].

$$RU_{s,j,g,h} \leq RUW_{j,g} \qquad \forall s \in \Omega_S, j \in \Omega_G, g \in \Omega_U, h \in \Omega_H \tag{27}$$

$$RD_{s,j,g,h} \leq RDW_{j,g} \qquad \forall s \in \Omega_S, j \in \Omega_G, g \in \Omega_U, h \in \Omega_H \tag{28}$$

$$(5\%) \times \left(\sum_{j\in\Omega_W} Wf_h.Pw_{s,j} + \sum_{j\in\Omega_{PV}} PVf_h.Pv_{s,j}\right) + (3\%) \times \left((1+LG_s)^s.\sum_{j\in\Omega_B} Lf_h.LD_j^{pk}\right) \leq$$
$$\sum_{j\in\Omega_G}\sum_{g\in\Omega_U} RU_{s,j,g,h} \qquad \forall s \in \Omega_S, h \in \Omega_H \qquad (29)$$

$$(5\%) \times \left(\sum_{j\in\Omega_W} Wf_h.Pw_{s,j} + \sum_{j\in\Omega_{PV}} PVf_h.Pv_{s,j}\right) + (3\%) \times \left((1+LG_s)^s.\sum_{j\in\Omega_B} Lf_h.LD_j^{pk}\right) \leq$$
$$\sum_{j\in\Omega_G}\sum_{g\in\Omega_U} RD_{s,j,g,h} \qquad \forall s \in \Omega_S, h \in \Omega_H \qquad (30)$$

The presented constraints of (31) to (34) guarantee that FRSR can be available at any time by ensuring that the upward and downward FRSR do not surpass the corresponding maximum FRSR capacity at $\tau$-minutes ($\tau$=10 minutes), and the maximum power capacity at the end of that hour. To this end, (31) and (32) confirm that the maximum FRSR capacity at $\tau$-minutes is satisfied for each TGU. Besides, (33) and (34) ensure the maximum power capacity for $\tau$-minutes and also at the end of the hour [30].

$$\tau/60 \times (P_{s,j,g,h} - P_{s,j,g,h-1}) + RU_{s,j,g,h} \leq RUW_{j,g} \qquad \forall s \in \Omega_S, j \in \Omega_G, g \in \Omega_U, h \in \Omega_H \qquad (31)$$

$$\tau/60 \times (P_{s,j,g,h-1} - P_{s,j,g,h}) + RD_{s,j,g,h} \leq RDW_{j,g} \qquad \forall s \in \Omega_S, j \in \Omega_G, g \in \Omega_U, h \in \Omega_H \qquad (32)$$

$$\tau/60 \cdot P_{s,j,g,h} + (1 - \tau/60) \cdot P_{s,j,g,h-1} + RU_{s,j,g,h} \leq [NE_{j,g} + N_{s,j,g}] \times Cap_{j,g}$$
$$\forall s \in \Omega_S, j \in \Omega_G, g \in \Omega_U, h \in \Omega_H \qquad (33)$$

$$RD_{s,j,g,h} \leq \tau/60 \cdot P_{s,j,g,h} + (1 - \tau/60) \cdot P_{s,j,g,h-1} \qquad \forall s \in \Omega_S, j \in \Omega_G, g \in \Omega_U, h \in \Omega_H \qquad (34)$$

According to the constraints of (35) to (39), the RPS policy is implemented. In (35) and (36), the wind and PV farms power capacity limits are formulated. The lower bound of wind and PV farms available units are assumed as a certain percentage of the total load demand energy at each planning stage. The mentioned percentage is chosen in a way that based on the RPS policy, the penetration of wind and PV farms at the last stage of the planning horizon reaches $\alpha$% of the total load demand energy, as expressed in (37). The constraints (38) and (39) guarantee that the installed wind and PV farms power capacity at each stage will remain until the end of planning horizon.

$$0 \leq Pw_{s,j} \leq Pw_j^{max} \qquad \forall s \in \Omega_S, j \in \Omega_W \qquad (35)$$

$$0 \leq Pv_{s,j} \leq Pv_j^{max} \qquad \forall s \in \Omega_S, j \in \Omega_{PV} \qquad (36)$$

$$[\alpha \times {}^s/_S] \times (1 + LG_s)^s . \sum_{h \in \Omega_H} \sum_{j \in \Omega_B} Lf_h . LD_j^{pk} \leq \sum_{h \in \Omega_H} (\sum_{j \in \Omega_W} Wf_h . Pw_{s,j} - PC_{s,j,h} + \qquad (37)$$

$$\sum_{j \in \Omega_{PV}} PVf_h . Pv_{s,j}) \qquad \forall s \in \Omega_S$$

$$Pw_{s-1,j} \leq Pw_{s,j} \qquad \forall s \in \Omega_S, j \in \Omega_W \qquad (38)$$

$$Pv_{s-1,j} \leq Pv_{s,j} \qquad \forall s \in \Omega_S, j \in \Omega_{PV} \qquad (39)$$

Wind curtailment, which is mainly due to wind power intermittency and transmission line congestion, should be minimized [12]. Therefore, especially when BES devices are utilized, it is needed to model the wind curtailment and load shedding (or ENS). In this regard, the constraints of (40) to (43) are introduced. The hourly wind curtailment is bounded by (40). In (41), the maximum allowable total wind curtailment at each stage of the planning horizon is defined as a certain percentage of the installed wind farms total energy at that stage. For load shedding or load curtailment modeling, the hourly bounds of permitted load shedding in each bus are specified using (42). Based on (43), the maximum permitted total load shedding in each stage is considered as a certain percentage of the total energy demand.

$$0 \leq PC_{s,j,h} \leq Wf_h . Pw_{s,j} \qquad \forall s \in \Omega_S, j \in \Omega_W, h \in \Omega_H \qquad (40)$$

$$\sum_{j \in \Omega_W} \sum_{h \in \Omega_H} PC_{s,j,h} \leq \beta \times \sum_{j \in \Omega_W} \sum_{h \in \Omega_H} Wf_h . Pw_{s,j} \qquad \forall s \in \Omega_S \qquad (41)$$

$$0 \leq LS_{s,j,h} \leq \gamma . (1 + LG_s)^s . Lf_h . LD_j^{pk} \qquad \forall s \in \Omega_S, j \in \Omega_B, h \in \Omega_H \qquad (42)$$

$$\sum_{j \in \Omega_B} \sum_{h \in \Omega_H} LS_{s,j,h} \leq \Phi \times (1 + LG_s)^s . \sum_{j \in \Omega_B} \sum_{h \in \Omega_H} Lf_h . LD_j^{pk} \quad \forall s \in \Omega_S \qquad (43)$$

To manage the intermittency of RESs output power, defer the construction of new transmission lines and TGUs, and reduce the wind and load curtailment, BES devices need to be incorporated in the power system expansion planning. To this end, the constraint of (44) to (50) are presented. In (44) and (45), the bounds of BES charging and discharging power are defined. The constraints of (46) and (47) determine the BES charging and discharging hourly state. The level of hourly stored energy in BES is considered as the summation of stored energy at the previous hour, and the exchanged energy at the given hour, as expressed in (48). In (49), the limits of BES stored energy level are defined. The constraint (50) guarantees that the installed BES in a given stage will remain until the end.

$$0 \leq \eta_c . Pc_{s,j,t,h} \leq CM_{j,t} . I_{s,j,t} \qquad \forall s \in \Omega_S, j \in \Omega_{BS}, t \in \Omega_T, h \in \Omega_H \qquad (44)$$

$$0 \leq {1}/{\eta_d} . Pd_{s,j,t,h} \leq CM_{j,t} . I_{s,j,t} \qquad \forall s \in \Omega_S, j \in \Omega_{BS}, t \in \Omega_T, h \in \Omega_H \qquad (45)$$

$$\eta_c . Pc_{s,j,t,h} \leq CM_{j,t} . U_{s,j,t,h} \qquad \forall s \in \Omega_S, j \in \Omega_{BS}, t \in \Omega_T, h \in \Omega_H \qquad (46)$$

$${1}/{\eta_d} . Pd_{s,j,t,h} \leq CM_{j,t} . (1 - U_{s,j,t,h}) \qquad \forall s \in \Omega_S, j \in \Omega_{BS}, t \in \Omega_T, h \in \Omega_H \qquad (47)$$

$$E_{s,j,t,h} = E_{s,j,t,h-1} + (\eta_c . Pc_{s,j,t,h}) - ({1}/{\eta_d} . Pd_{s,j,t,h}) \qquad \forall s \in \Omega_S, j \in \Omega_{BS}, t \in \Omega_T, h \in \Omega_H \qquad (48)$$

$$0 \leq E_{s,j,t,h} \leq SM_{j,t} . I_{s,j,t} \qquad \forall s \in \Omega_S, j \in \Omega_{BS}, t \in \Omega_T, h \in \Omega_H \qquad (49)$$

$$I_{s-1,j,t} \leq I_{s,j,t} \qquad \forall s \in \Omega_S, j \in \Omega_{BS}, t \in \Omega_T \qquad (50)$$

The constraint given in (51) determines the power flow limits of existing lines. In (52), the hourly power flows of existing lines is determined.

$$-Pe_l^{max} \leq Pe_{s,l,h} \leq Pe_l^{max} \qquad \forall s \in \Omega_S, l \in \Omega_{el}, h \in \Omega_H \qquad (51)$$

$$Pe_{s,l,h} - \sum_{j \in \Omega_B} \Psi . B_l . A_j^l . \theta_{s,j,h} = 0 \qquad \forall s \in \Omega_S, l \in \Omega_{el}, h \in \Omega_H \qquad (52)$$

The power flow limits of new constructed lines are defined in (53). In (54), the power flow of new constructed lines is defined using a big-M value. The constraint of (55) confirms that the constructed line at a given stage will remain at the next stages.

$$-Pl_l^{max} . Y_{s,l,c} \leq Pl_{s,l,c,h} \leq Pl_l^{max} . Y_{s,l,c} \qquad \forall s \in \Omega_S, l \in \Omega_{nl}, c \in \Omega_C, h \in \Omega_H \qquad (53)$$

$$-\mathcal{M}_l . (1 - Y_{s,l,c}) \leq Pl_{s,l,c,h} - \sum_{j \in \Omega_B} \Psi . B_l . K_j^l . \theta_{s,j,h} \leq \mathcal{M}_l . (1 - Y_{s,l,c})$$
$$\forall s \in \Omega_S, l \in \Omega_{nl}, c \in \Omega_C, h \in \Omega_H \qquad (54)$$

$$Y_{s-1,l,c} \leq Y_{s,l,c} \qquad \forall s \in \Omega_S, l \in \Omega_{nl}, c \in \Omega_C \qquad (55)$$

In (56), at each stage, the hourly nodal power balance including the TGUs and RESs output power considering wind curtailment, BES devices power exchange load shedding is formulated. In Fig. 2, all mentioned possible power exchange options in nodal power balance, are illustrated.

$$\sum_{g\in\Omega_U} P_{s,j,g,h} + [(Wf_h.Pw_{s,j}) - PC_{s,j,h} + (PVf_h.Pv_{s,j})] + \left[\sum_{t\in\Omega_T}(Pd_{s,j,t,h} - Pc_{s,j,t,h})\right]$$

$$-\sum_{l\in\Omega_{el}} A_j^l.Pe_{s,l,h} - \sum_{l\in\Omega_{nl}}\sum_{c\in\Omega_C} K_j^l.Pl_{s,l,c,h} = \left((1+LG_s)^s.Lf_h.LD_j^{pk}\right) - LS_{s,j,h} \quad (56)$$

$$\forall s \in \Omega_S, j \in \Omega_B, h \in \Omega_H$$

## 2. Benders Dual Decomposition

The proposed multi-stage TGSEP model is solved using an accelerated BDD algorithm. In this regard, the presented MILP formulation is reformulated using an equivalent compact form. Then, by introducing a Master Problem (MP), and two Dual Sub-Problems (DSPs), the main problem is decomposed. In MP, the binary decision variables are optimized, and then they are assumed as constant parameters in DSPs. In DSPs, the continuous variables are optimized, and also the feasibility or optimality of MP solution is examined and then the required cuts are constructed and transferred to the MP. The equivalent compact form of all equations in (1) to (56) is presented in (57) to (61). The objective function presented by (1) to (15) is compacted by (57). The constraints given in (21), (50), and (55) are recast by (58). The constraint of (59) refers to the equality constraints in (17), (20), (48), and (52). The constraint of (60) corresponds to (16), (18), (19), (22) to (47), (49), (51), (53), and (54). The nodal power balance constraint of (56) is compacted by (61). The compact dual variables of (59), (60), and (61) are introduced as $\lambda, \mu$, and $\sigma$, respectively. The vectors of $I$, and $I_R$ stand for investment costs, and $O_P$ is the operation cost vector.

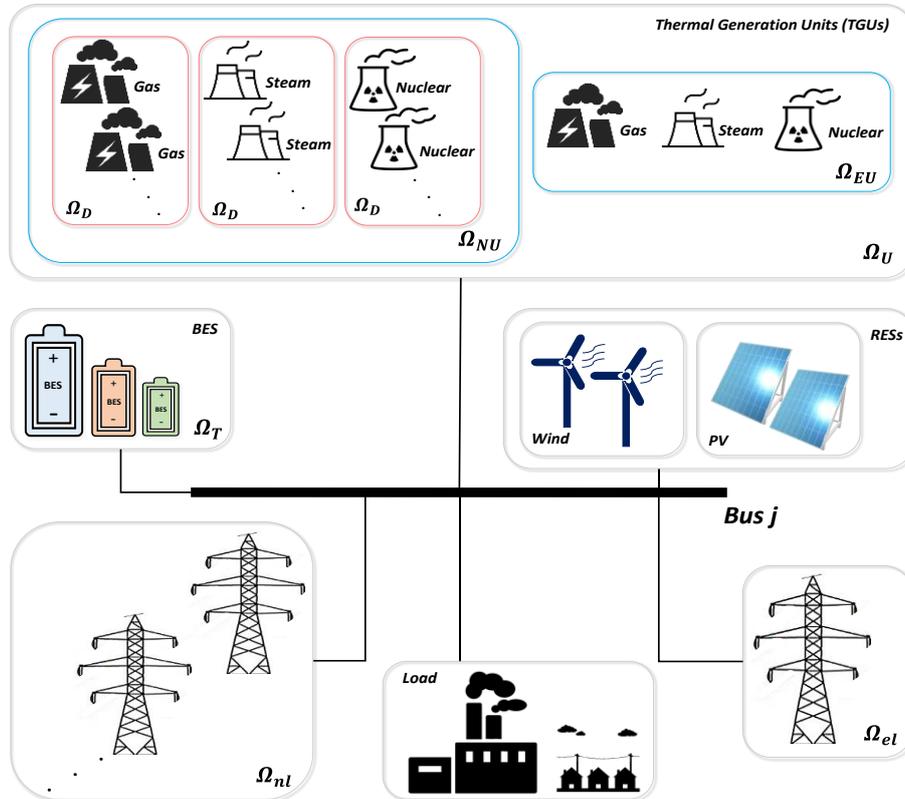

Fig. 2. All possible power exchange options in nodal power balance

The coefficient vectors of $A, B, C_1, C_2, D_1, D_2, E_1, E_2, G_1, G_2, H_1, H_2, J, K, L$, and $M$, are all the relevant matrices of the general framework of formulations, and $T$ is the transpose operator.

$$\text{Min} \quad I^T Y + I_R^T R + O_P^T P \tag{57}$$

s.t.

$$AY \geq B \tag{58}$$

$$C_1 Y + D_1 R + E_1 P + G_1 F = H_1 \quad : \lambda \tag{59}$$

$$C_2 Y + D_2 R + E_2 P + G_2 F \geq H_2 \quad : \mu \tag{60}$$

$$JR + KP + LF = M \quad : \sigma \tag{61}$$

$Y \in \{0,1\}, \quad R \ \& \ P \geq 0, \quad F: free$

$Y = \{Y, X, I, U\}, \quad F = \{Pe, Pl, \theta\}$

$P = \{P, Ps, N, RU, RD, Pd, Pc, E, PC, LS\}, \quad R = \{Pw, Pv\}$

$\sigma \ \& \ \lambda: free, \quad \mu \geq 0$

- **Master Problem**

Before introducing MP formulations, it should be noted that $\pi$ is the compact dual variable of (62) as the SP auxiliary constraint.

$$I_{SP} Y_{SP} = \overline{Y} \quad : \pi \tag{62}$$

$I_{SP}$: Identity Matrix, $\pi$: free

The formulation of MP is introduced in (63) to (66). The objective function of MP that is equal to the Lower Bound (LB) of the problem is presented in (63). The constraint of (64) determines the minimum value of LB when the DSP solution is unbounded. In (65) and (66), the optimality and feasibility cuts are defined. Note that $v$ is the BDD algorithm iteration number.

$$Min\ Z_{LB} \qquad (63)$$
s.t.
$$Z_{LB} \geq I^T Y \qquad (64)$$

$$Z_{LB} \geq I^T Y + [M^T\overline{\sigma} + H_1^T\overline{\lambda} + H_2^T\overline{\mu}]^{(v)} + \overline{\pi}^{(v)}.(Y - \overline{Y}^{(v-1)}) \qquad (65)$$

$$[M^T\overline{\sigma} + H_1^T\overline{\lambda} + H_2^T\overline{\mu}]^{(v)} + \overline{\pi}^{(v)}.(Y - \overline{Y}^{(v-1)}) \leq 0 \qquad (66)$$

& (58)

- **Dual Sub-Problem**

The DSP is presented by (67) to (71). According to the solution of MP, the binary decision variables (i.e., $\overline{Y}$) are obtained and fixed for DSP. Then, the Linear Programming (LP) formulation of DSP is solved.

$$Max\ H_1^T\lambda + H_2^T\mu + M^T\sigma + \overline{Y}^T\pi \qquad (67)$$
s.t.
$$E_1^T\lambda + E_2^T\mu + K^T\sigma \leq O_C \qquad (68)$$

$$D_1^T\lambda + D_2^T\mu + J^T\sigma \leq I_R \qquad (69)$$

$$G_1^T\lambda + G_2^T\mu + L^T\sigma = 0 \qquad (70)$$

$$C_1^T\lambda + C_2^T\mu + I_{SP}^T\pi \leq 0 \qquad (71)$$

After solving DSP, if the solution is bounded, the Upper Bound (UB) is calculated using (72), and based on (65), the optimality cut is generated.

$$UB = H_1^T\lambda + H_2^T\mu + M^T\sigma + \overline{Y}^T\pi + I^T\overline{Y} \qquad (72)$$

If the solution is unbounded, the Modified DSP (MDSP) is solved to generate the feasibility cut of (66).

- **Modified DSP**

The objective function of MDSP is equal to (67), and its constraints are similar to (68) to (71) except that the right-hand sides of them are assigned to zero. Also, the MDSP formulation contains the auxiliary constraint of (73).

$$\sigma \leq 1 \qquad (73)$$

Finally, at the end of the BDD algorithm, if the tolerance defined in (74) is satisfied,

the algorithm is terminated, and otherwise, the next iteration is started.

$$\frac{(UB-LB)}{UB} \leq \tau \tag{74}$$

- **Acceleration Tool**

The proposed BDD algorithm is accelerated by utilizing the pareto optimality cut, which can reduce the number of algorithm iterations [35]. The POC is a dominant optimality cut. After obtaining a bounded solution for DSP, the POC, as a dominant optimality cut, is generated by solving the second form of DSP (SDSP). According to (75), in the objective function of SDSP, the $\overline{Y}^T$ is replaced by $\widehat{Y}^{T(v)}$, as the vector of $Y$ core points in iteration $v$.

$$\widehat{Y}^{T(v)} = 1/2 \left( \widehat{Y}^{T(v-1)} + \overline{Y}^T \right) \tag{75}$$

In the proposed SDSP, in addition to (68) to (71), the given constraint in (76) is also considered. The parameter of $\bar{Z}_{DSP}$ in (76) is equal to the value of the DSP objective function (i.e., (67)). Finally, the optimality cut of (65) is generated by solving the SDSP.

$$H_1^T \boldsymbol{\lambda} + H_2^T \boldsymbol{\mu} + M^T \boldsymbol{\sigma} + \overline{Y}^T \boldsymbol{\pi} = \bar{Z}_{DSP} \tag{76}$$

### III. OVERALL STRUCTURE OF THE PROPOSED MODEL

The overall structure of the proposed TGSEP model based on the BDD formulation is depicted in Fig. 3. As illustrated in Fig. 3, all the investment, operation, maintenance, and emission required input data of the TGUs, RESs, transmission lines, and BES devices are prepared. In each iteration of the BDD algorithm, the MP is solved to obtain the construction cost of new TGUs, lines, and BES devices, along with the BES devices hourly charging/discharging states. Then, after calculating the LB and fixing the obtained binary variable values, the DSP is solved to determine the total installed capacity of RESs, hourly power output of existing and new constructed TGUs and RESs farms, transmitted power flow of all lines, energy exchanges of BES devices, the upward and downward FRSR requirement, the load shedding, and the wind curtailment. Regarding the DSP solution, if it is unbounded, to modify the MP solution and remove the related extreme rays, the feasibility cut is generated by solving the MDSP, otherwise, if the DSP solution is bounded, after calculating the UB, the POC is generated for the MP by solving the SDSP. Meanwhile, the termination criterion (i.e., (74)) is checked at the end of the BDD algorithm, and if it is satisfied, the process will be ended; otherwise, the next iteration will be started.

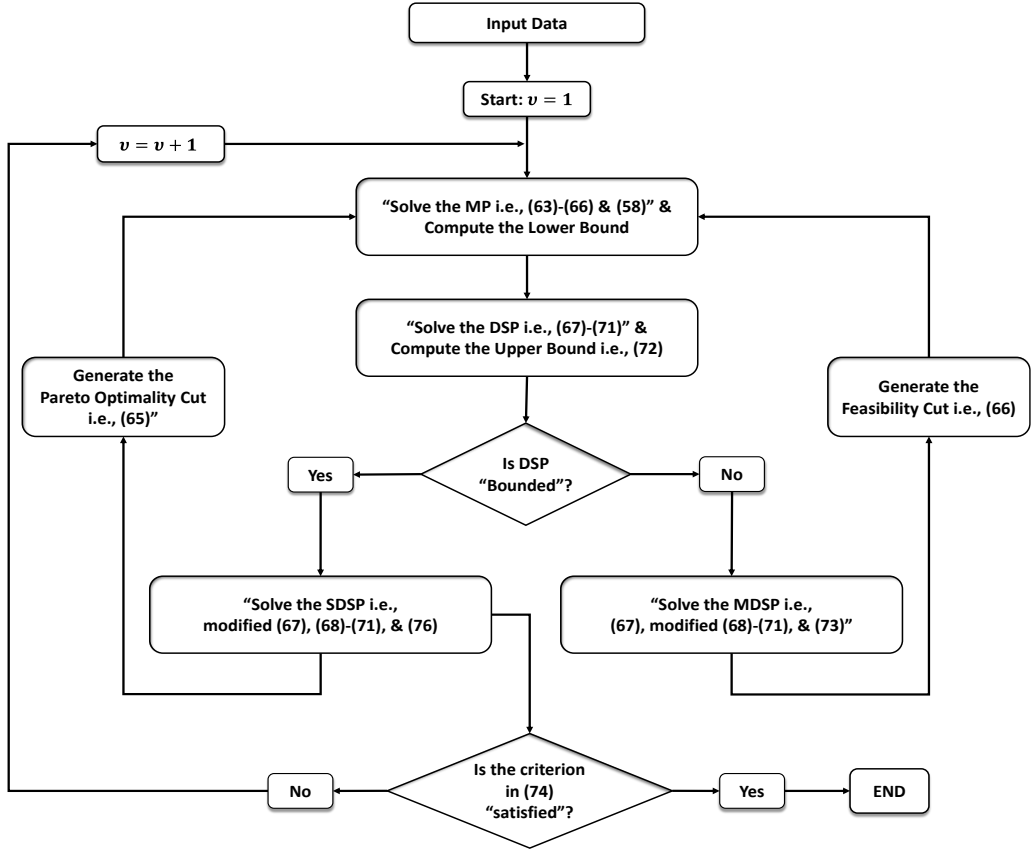

Fig. 3. The overall structure of the proposed TGSEP model

## IV. REPRESENTATIVE HOURS

In order to capture the short-term uncertainties of load demand and RESs output power, and reduce the problem computational burden, an effective hierarchical clustering algorithm is developed. Historical load demand and RESs output power data of Norway in 2020 are utilized [36]. Note that although the weather data including wind speed and solar irradiances are input data for RESs generation, in the proposed model weather data are considered indirectly within RESs output power data. Although most clustering algorithms focus on reducing the computational complexity, the hierarchical clustering algorithm proposed in [13] can capture the dynamic and operation of RESs more accurately, and preserve the chronology of input time series throughout all stages of the planning horizon. As shown in Fig. 4, based on the algorithm presented in [13], a procedure is added to reduce the input time series data burden to minimize the number of similar days. Also, the metrics presented in [37] are used for achieving the minimum proper number of representatives. Therefore, a total number of 96 representative hours are extracted. In Fig. 4, the extracted 96 representative hours are presented throughout a year considering each representative weight (or the probability of each representative occurrence).

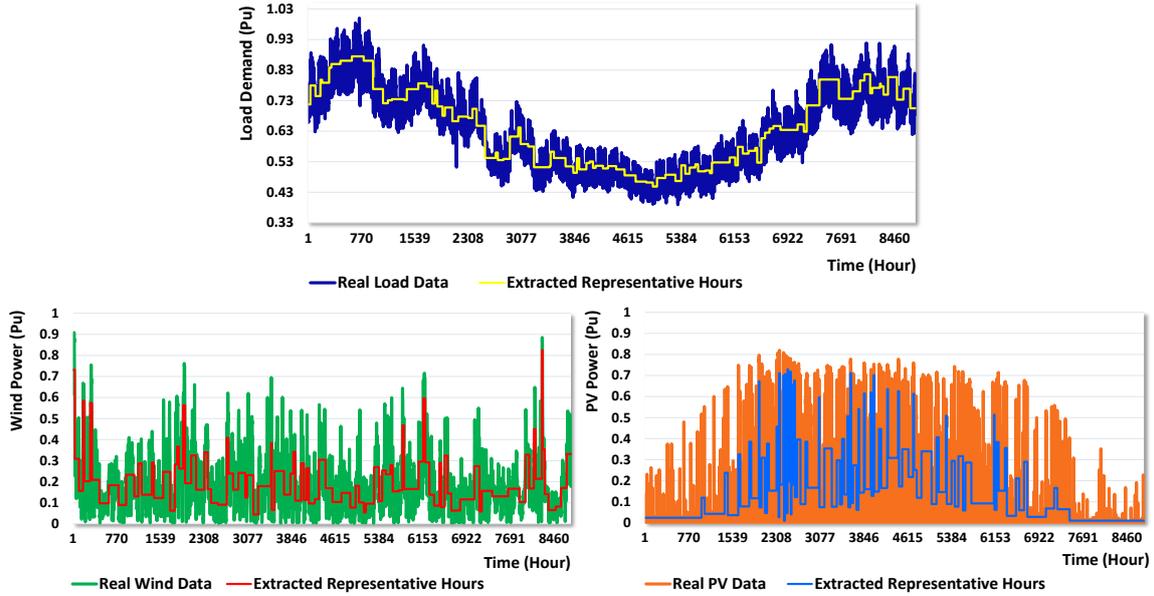

Fig. 4. The extracted representative hours by the utilized hierarchical clustering method

## V. NUMERICAL RESULTS

To validate the effectiveness of the proposed TGSEP model, the IEEE RTS 24-Bus test system is utilized. All the basic data of this system are extracted from [38] and [39]. All the related costs of existing and candidate TGUs and RESs are revised according to [40], and the related costs of BES devices are taken from [27]. Moreover, all input data, including existing and candidate TGUs, RESs, transmission lines, and BES devices data, along with peak load data in buses, and all costs, e.g., investment, operational, maintenance, and CO2 emission costs, are given in [41]. The planning horizon is 14 years, which is considered as 7 stages with a 2-year duration. The $CL$, $Rt$, and $SS$ for single (double) circuit corridors are considered as 1 (1.6), 0.034 (0.039) $10^6\$/Km$, and 3.358 (6.38) $10^6\$$, respectively [42]. The $\eta_c$ and $\eta_d$ are 0.9 [27], and $Cd$ is assumed to be 5 $\$/Mwh$ [43]. The $CLS$ and $CWC$ are both assumed as 1000 $\$/MWh$ [31]. The $\alpha$, $\beta$, and $TRR^{min}$ are assumed 10%, 50%, and 15%, respectively. To investigate the impacts of load shedding, the value of $\gamma$ is assumed to be in the range of 0 to 10%, and $\Phi$ is in the range of 0 to 3%. The $LG$ and $i$ are equal to 10%, and 5%, respectively, and $\Psi$ is 100 $MVA$. Note that the load curve is proportionally scaled by the $LG$ over the planning stages.

Three types of BES devices are considered as candidates to be installed in existing buses 10, and 18, along with new buses 25, and 26. Five buses, i.e., 3, 7, 23, 25, and 26, are assumed as candidates for installing wind farms. Buses 1, 22, and 26, are assumed as PV farms candidate buses. Two types for each Gas Peaking, Steam, and Nuclear TGU technologies are considered as candidate units to be constructed in buses 7, 18, and 23. The new expansion buses 25 and 26 can be connected to the system through new double-circuit corridors of 21-25, and 16-26. Eleven new lines are considered as transmission expansion candidates. As mentioned, all the required data are presented in [41].

Based on Tables I & II, to confirm the proposed TGSEP model effectiveness, the simulation results are presented in five different cases. In case 1, the proposed TGSEP model is simulated considering the load shedding possibility with $\gamma = 10\%$, and $\Phi = 3\%$. For other cases, i.e., cases 2, 3, 4, and 5, both $\gamma$ and $\Phi$ are assumed to be zero. In case 2, both Low-Carbon Policy (LCP) and BES devices impacts are ignored. In case 3, the simulation is conducted without LCP, but BES devices are taken into account. Case 4 is similar to case 3, except that LCP is considered, while the BES devices impacts are not

Table I. The simulation results of the proposed TGSEP model considering load shedding possibility

| Case | Description | | Stage | New Line (From-To) | New TGU (MW) | New RES | | New BES (MW-MWh) |
|---|---|---|---|---|---|---|---|---|
| | | | | | | Wind (MW) | PV (MW) | |
| 1 | BES | ✓ | 1 | — | **18**[3]: Gas 160, **23**: Gas 320 | **3**: 200 **23**: 39.2 | — | **18**: (100-600) & (200-1200) |
| | | | 2 | — | **18**: ST[4] 320, **23**: Gas 160 | **23**: 43.3 | — | **18**: (50-300) |
| | | | 3 | 7-8 | — | **23**: 90.15 **7**: 200 | — | — |
| | LCP | ✓ | 4 | — | **18**: Gas 320, **23**: ST 320 | **23**: 27.35 | **1**: 31 **22**: 100 | **10**: (50-300) |
| | | | 5 | (21-25)[2] & (16-26)[2] | — | **25**: 89.4 **26**: 181.85 | — | **25**: (50-300) |
| | | | 6 | 6-10 & 7-8 | **7**: Gas 160 & ST 320 | **25**: 116.35 **26**: 20 | **1**: 37.6 **26**: 200 | **26**: (50-300) |
| | LSP[1] | ✓ | 7 | (7-8)[2] & (16-26)[2] | **7**: Gas 320, **18**: ST 320 | **25**: 139.25 **26**: 198.15 | **1**: 31.4 | **25**: (100-600) |
| | | | $C_{Inv}^5$: | 1.2052 | 15.0018 | 7.729 | 0.8659 | 3.155 |
| | | | $TC_{Inv}$: 27.957 | | $TC_O$: 62.475 | $TC_M$: 23.567 | $TC_E$: 5.324 | $TC_P$: 119.323 |

**1**: Load Shedding Possibility (LSP), **2**: Double-circuit line, **3**: Bus number, **4**: Steam, **5**: All presented costs are $\times 10^8$ \$ incorporated in the model. In case 5, all the expansion planning and operational tools are considered. The complete results of case 1, and cases 2 to 5 are presented in Tables I & II. For better illustration, the presented results of case 5 are depicted in Fig. 5.

According to Table I, by considering the load shedding possibility in case 1, the total energy demand of 62.771 $GWh$ will not be supplied during the planning horizon. In this case, 3 single-circuit and 4 double-circuit lines, along with 10 new TGUs are constructed and the total wind curtailment is zero. The $TC_{Inv}, TC_O, TC_M, TC_E$, and $TC_P$ are obtained as 27.957, 62.475, 23.567, 5.324, and 119.323 $10^8$ \$, respectively. In case 1, the total CO2 emission is 43.091 Million-ton CO2 during the planning horizon. In case 2, as presented in Table II, when the load shedding possibility, and both LCP and BES devices are ignored, 3 single-circuit and 5 double-circuit transmission lines, along with 12 new TGUs are constructed. The total amount of wind curtailment, and the total emission are as 0.676 $GWh$ and 46.853 Million-ton CO2, respectively. In case 2, 3.762 Million-ton CO2 more than case 1 is emitted during the planning horizon as illustrated in Fig. 7. In this case, the $TC_P$ is 0.807 $10^8$ \$ more expensive than case 1. The noticeable amount of wind curtailment in case 2 is due to ignoring BES devices impacts. Based on case 3, when the load shedding possibility and LCP are ignored, 3 single-circuit and 5 double-circuit lines,

Table II. The simulation results of the proposed TGSEP model ignoring load shedding possibility

| Case | Description | | Stage | New Line (From-To) | New TGU (MW) | New RES Wind (MW) | New RES PV (MW) | New BES (MW-MWh) |
|---|---|---|---|---|---|---|---|---|
| 2 | BES | ✗ | 1 | — | **18**: Gas 320, **23**: Gas 160, Gas 320 & ST 320 | **3**: 100.25, **7**: 28 | — | — |
| | | | 2 | — | — | **3**: 88.5, **23**: 200 | — | — |
| | | | 3 | 7-8 | **18**: Gas 160 & ST 320 | **3**: 11.25, **7**: 49.7 | — | — |
| | LCP | ✗ | 4 | (21-25)[1] & (16-26)[1] | — | **7**: 28.3, **25**: 206, **26**: 181.85 | — | — |
| | | | 5 | (7-8)[1] | **7**: Gas 160 & ST 320 | **7**: 94, **25**: 26.4, **26**: 22 | — | — |
| | | | 6 | 6-10, 8-10 & (21-25)[1] | **7**: Gas 320, **18**: ST 320 | **25**: 159 | **1**: 89 | — |
| | LSP | ✗ | 7 | (16-26)[1] | **18**: ST 440, **23**: ST 440 | **25**: 8.6, **26**: 164.3 | **1**: 11, **22**: 100 **26**: 167.96 | — |
| | | | $C_{Inv}$: | 1.897 | 23.233 | 8.627 | 0.3538 | 0 |
| | | | $TC_{Inv}$: 34.1108 | $TC_O$: 60.4127 | $TC_M$: 25.607 | $TC_E$: 0 | | $TC_P$: 120.13 |
| 3 | BES | ✓ | 1 | — | **18**: Gas 160 & Gas 320, **23**: Gas 320 & ST 320 | **3**: 128.25 | — | **10**: (100-600), **18**: (100-600) |
| | | | 2 | — | — | **3**: 71.25, **7**: 28, **23**: 188.75 | — | **10**: (50-300), **18**: (50-300) |
| | | | 3 | 7-8 | **18**: ST 320, **23**: Gas 160 | **7**: 61 | — | — |
| | LCP | ✗ | 4 | (21-25)[1] & (16-26)[1] | — | **7**: 111, **23**: 11.25 **25**: 232.4, **26**: 61.5 | — | **25**: (50-300) |
| | | | 5 | (7-8)[1] | **7**: Gas 160 & ST 320 | **26**: 142.25 | — | — |
| | | | 6 | 6-10, 8-10 & (21-25)[1] | **7**: Gas 320, **18**: ST 320 | **25**: 159 | **1**: 89 | — |
| | LSP | ✗ | 7 | (16-26)[1] | **18**: ST 440, **23**: ST 440 | **25**: 8.6, **26**: 164.4 | **1**: 11, **22**: 100 **26**: 167.96 | — |
| | | | $C_{Inv}$: | 1.897 | 23.233 | 8.627 | 0.3538 | 2.388 |
| | | | $TC_{Inv}$: 36.4988 | $TC_O$: 57.6603 | $TC_M$: 25.607 | $TC_E$: 0 | | $TC_P$: 119.766 |
| 4 | BES | ✗ | 1 | — | **18**: Gas 320, **23**: Gas 160, Gas 320 & ST 320 | **3**: 100.3, **7**: 28 | — | — |
| | | | 2 | — | — | **3**: 88.45, **23**: 200 | — | — |
| | | | 3 | 7-8 | **7**: ST 320, **18**: Gas 160 | **3**: 11.25, **7**: 49.65 | — | — |
| | LCP | ✓ | 4 | 2×(21-25)[1] | — | **7**: 16, **25**: 400 | — | — |
| | | | 5 | (16-26)[1] | **7**: Gas 160, **18**: ST 320 | **26**: 142.3 | — | — |
| | | | 6 | 6-10, 8-10 & (7-8)[1] | **7**: Gas 320, **18**: ST 320 | **7**: 106.35, **26**: 52.85 | **22**: 88.6 | — |
| | LSP | ✗ | 7 | (16-26)[1] | **18**: ST 440 **23**: ST 440 | **26**: 173 | **1**: 100, **22**: 11.4 **26**: 167.96 | — |
| | | | $C_{Inv}$: | 1.9081 | 23.233 | 8.627 | 0.3533 | 0 |
| | | | $TC_{Inv}$: 34.1214 | $TC_O$: 60.544 | $TC_M$: 25.607 | $TC_E$: 5.617 | | $TC_P$: 125.889 |
| 5 | BES | ✓ | 1 | — | **18**: Gas 320 & ST 320 **23**: Gas 160 & Gas 320 | **3**: 128.25 | — | **10**: (100-600), **18**: (50-300) & (200-1200) |
| | | | 2 | — | — | **3**: 71.25, **7**: 18, **23**: 198.75 | — | **10**: (50-300) |
| | | | 3 | 7-8 & (21-25)[1] | — | **7**: 182, **25**: 195.5 | — | **25**: (50-300) |
| | LCP | ✓ | 4 | (16-26)[1] | **18**: Gas 160, **23**: ST 320 | **25**: 37, **26**: 62.75 | — | **26**: (50-300) |
| | | | 5 | (7-8)[1] | **7**: Gas 160 & ST 320 | **26**: 142.3 | — | — |
| | | | 6 | 6-10, 8-10 & (21-25)[1] | **7**: Gas 320, **23**: ST 320 | **25**: 159 | **22**: 87.8 | — |
| | LSP | ✗ | 7 | (16-26)[1] | **18**: ST 440, **23**: ST 440 | **25**: 8.5, **26**: 164.4 | **1**: 100, **22**: 12.2 **26**: 167.7 | — |
| | | | $C_{Inv}$: | 2.052 | 22.216 | 9.224 | 0.3525 | 3.4939 |
| | | | $TC_{Inv}$: 37.338 | $TC_O$: 57.01 | $TC_M$: 25.572 | $TC_E$: 5.277 | | $TC_P$: 125.197 |

[1]: Double-circuit line

and 12 new TGUs are constructed. Although, in comparison to case 2, the $TC_{Inv}$ indicates

2.388 $10^8$ $ more cost in case 3, the $TC_O$, and $TC_P$ are 2.7524 $10^8$ $, and 0.364 $10^8$ $ less expensive than case 2, as illustrated using LCOE in Fig. 6. This cost reduction in case 3 is due to considering BES devices. Comparing cases 3 and 1 shows an increase of 8.5418, and 0.443 $10^8$ $ in $TC_{Inv}$ and $TC_P$ in case 3, which is due to the lack of load shedding. In case 3, the total amount of wind curtailment is 0.438 $GWh$. It is because by ignoring the LCP, the share of TGUs in load supplying can be increased without any limitation. The total amount of wind curtailment in case 3 is 0.238 $GWh$ less than case 2 that confirm the efficiency of using BES. In this case, 2.36 Million-ton CO2 more than case 1, and 1.402 Million-ton CO2 less than case 2, is emitted during the planning horizon as illustrated in Fig. 7, and Fig. 8. By ignoring the BES devices in case 4, 3 single-circuit and 5 double-circuit lines, along with 12 new TGUs are constructed. In this case, the total wind curtailment, and CO2 emission are 0.161 $GWh$, and 46.484 Million-ton CO2, which both are occurred due to the lack of BES devices. The CO2 emission in case 4 is 3.393 Million-

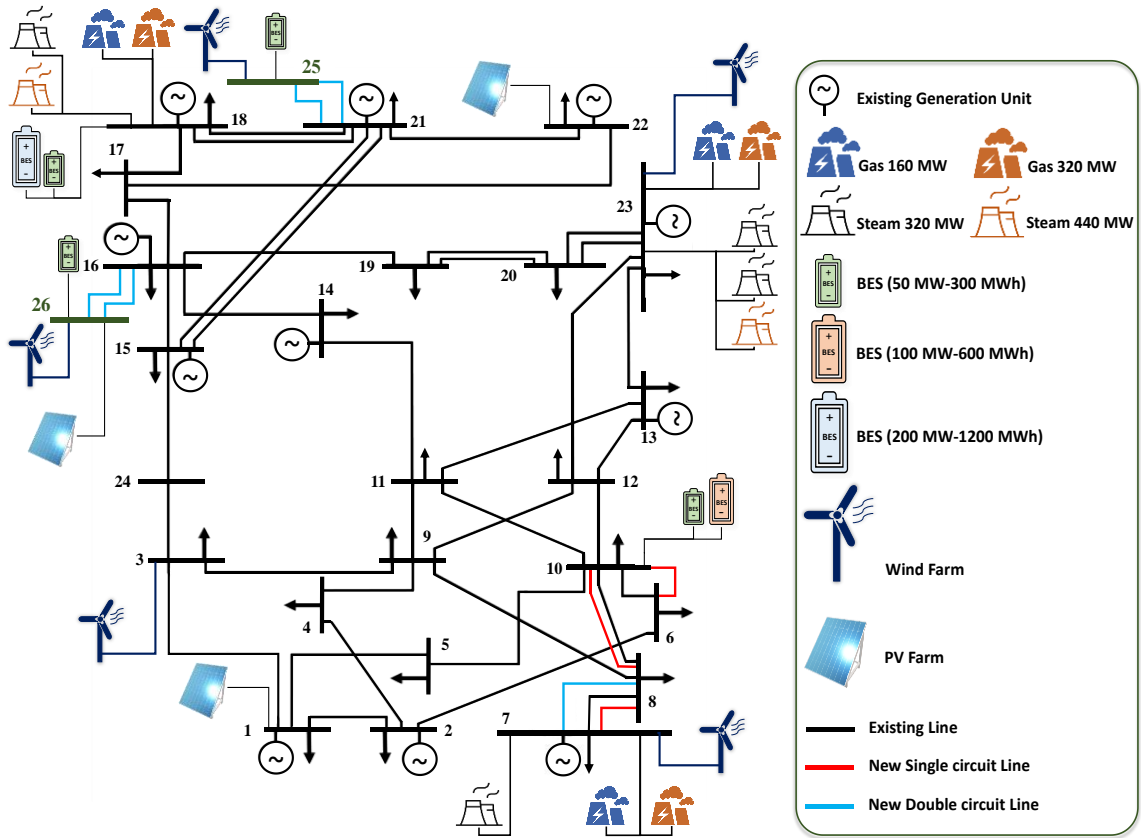

Fig. 5. The illustration of the obtained expansion plan of Case 5

ton, and 1.033 Million-ton more than cases 1 and 3, respectively. Also, due to considering the LCP, the CO2 emission in case 4 is 0.369 Million-ton less than cases 2. As shown in Fig. 7, and Fig. 8, although in case 4 the LCP is considered, after case 2, this case results in the worst-case CO2 emission. The $TC_P$ of case 4 is 6.566 $10^8$ \$, 5.759 $10^8$ \$, and 6.123 $10^8$ \$ more expensive than cases 1, 2, and 3, respectively. In the last case, i.e., case 5, both the LCP and BES devices are considered and 3 single-circuit and 5 double-circuit lines, and 12 new TGUs are constructed. The $TC_P$ of case 5 is 125.197 $10^8$ \$, that shows 0.692 $10^8$ \$ cost saving compared to case 4 due to BES devices utilization, as shown in Fig. 6. Although the $TC_P$ of case 5 is 5.431 more expensive than case 3, the total CO2 emission of case 5 is 44.003 Million-ton that is 1.448 Million-ton CO2 less than case 3, as illustrated in Fig. 7. Also, the total CO2 emission of case 5 is 2.85 Million-ton less than case 2. It should be noted that, although LCP tries to reduce the CO2 emission over time, in power systems with high shares of TGUs, its main focus is on reducing the CO2 emission growth rate. Regarding LCP and BES devices, in case 5 the total wind curtailment reaches to zero. According to the presented results in Table II and Fig. 5, the new constructed lines in case 5 are in corridors 7-8, 6-10, and 8-10, and new buses 25 and 26 are connected to the system through 4 double-circuit lines in new corridors 21-25 and 16-26, respectively. Also, a new double-circuit line is constructed in corridor 7-8. In bus 7, 2 new Gas and one new Steam

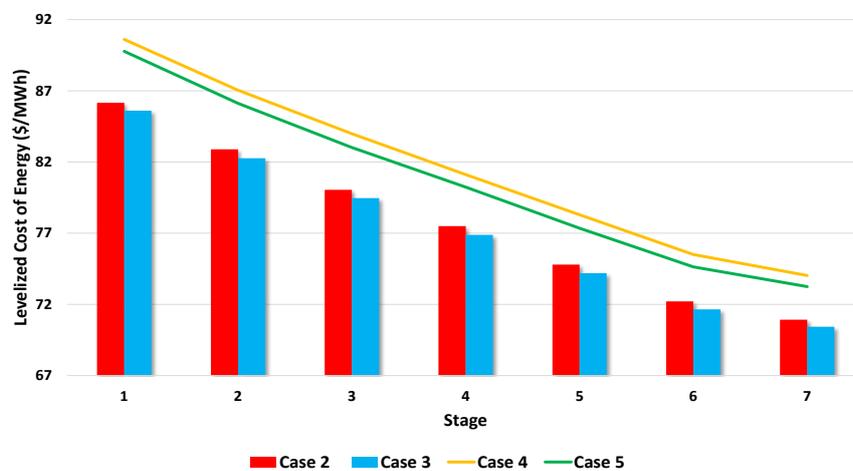

Fig. 6. The cumulative total planning cost per generated energy of cases 2 to 5 over the planning horizon

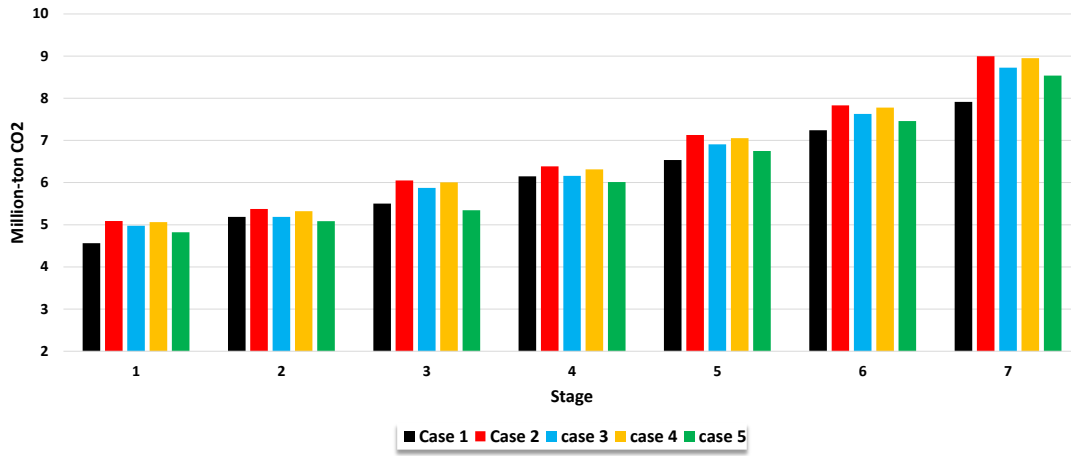

Fig. 7. The total CO2 emission of all Cases over the planning horizon

generation units, and a wind farm are constructed. Two new Gas and 2 new Steam generation units, and 2 BES devices in bus 18, along with 2 new Gas and 3 new Steam generation units, and also a wind farm in bus 23 are constructed, as shown in Fig 5. Additionally, 3 wind farms are installed in existing bus 3, and new buses 25 and 26. The PV farms are installed in existing buses 1 and 22, and new bus 26. Two BES devices are installed in bus 10, also buses 25 and 26 are each equipped with one BES device. For comparing the obtained results of cases 2 to 5, the cumulative total planning cost per generated power in $/MWh over the planning horizon, as Levelized Cost of Energy (LCOE) is illustrated in Fig. 6. In order to compare the $CO_2$ emission of all cases 1 to 5

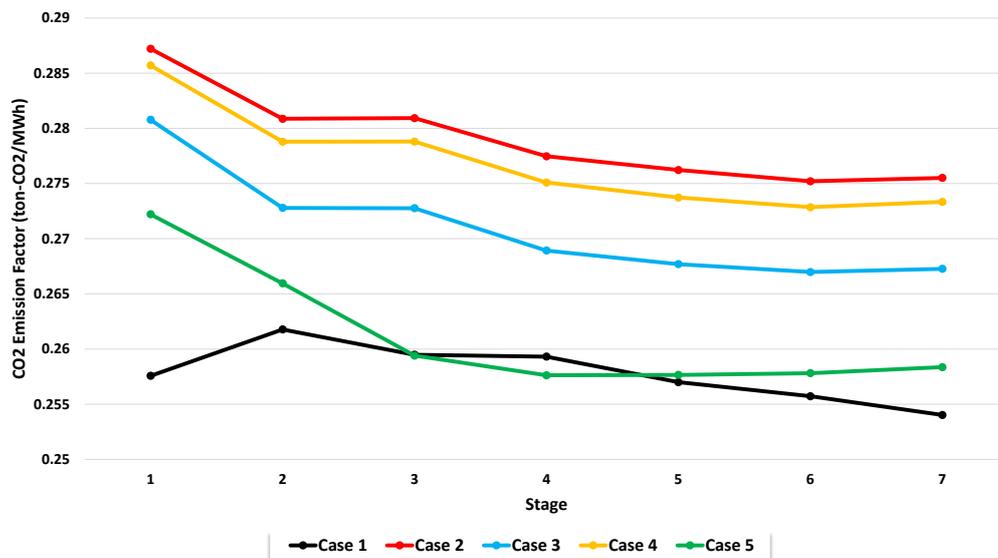

Fig. 8. The cumulative $CO_2$ emission factor in ton-$CO_2$/MWh for all Cases over the planning horizon

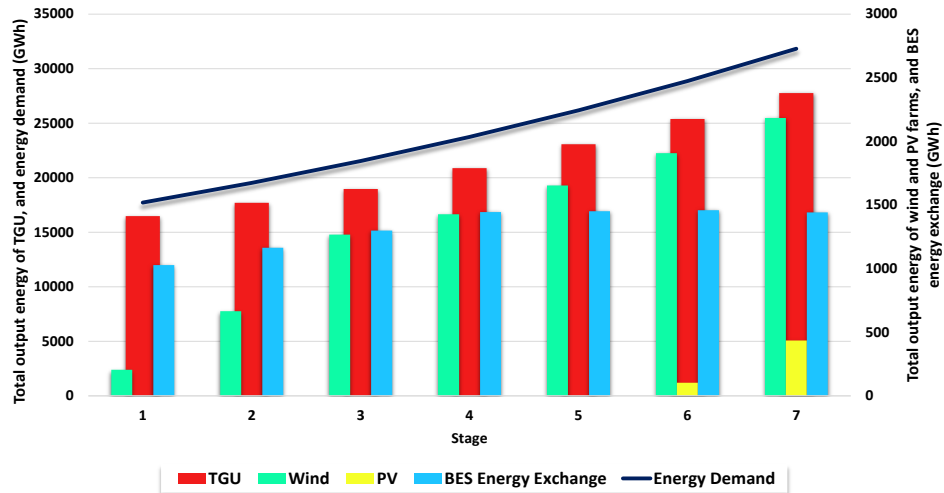

Fig. 9. The total share of all units in supplying energy demand at each stage of Case 5

during all planning stages, Fig. 8 is presented. In Fig. 8, the cumulative CO2 emission is calculated in ton-CO2/MWh. It should be noted that in the proposed model for all cases the RPS policy is considered. According to Fig. 8, case 2 in which both LCP and BES devices are ignored owns the worst CO2 emission factor in all stages. The influence of BES devices and LCP on reducing the CO2 emission can be inferred by comparing case 2, with cases 3, and 4, respectively. As shown in Fig. 8, cases 1 and 5 result in minimum CO2 emission factor in all stages. Note that case 1 is similar to case 5, except that in this case load shedding possibility is considered. Note that, as it can be seen in Fig. 8, although the overall trend in emitted CO2 reduction is descending, at some planning stages such as the last stage of cases 2 to 5, the cumulative total emitted CO2 per generated energy in ton-CO2/MWh is slightly increased. It is because the objective function of the proposed model, as shown in Fig. 1, includes different terms such as investment cost of new devices, operation cost, maintenance cost, and the emission cost. Therefore, the proposed model tries to minimize the total costs, not just the CO2 emission cost. In Fig. 9, for each planning stage of case 5, the total share of output energy of TGUs, wind and PV farms, and also BES devices energy exchange in supplying the total energy demand are illustrated to confirm the impact of RPS policy in long term horizon. In order to confirm the operational flexibility of proposed model, the hourly output energy of all units, including TGU, wind, and PV farms, along with hourly energy demand and total FRSR at the last planning stage of case 5 are presented in Fig. 10. The key role of BES devices in energy demand supplying and reducing RES curtailment is clearly shown in Fig. 10.

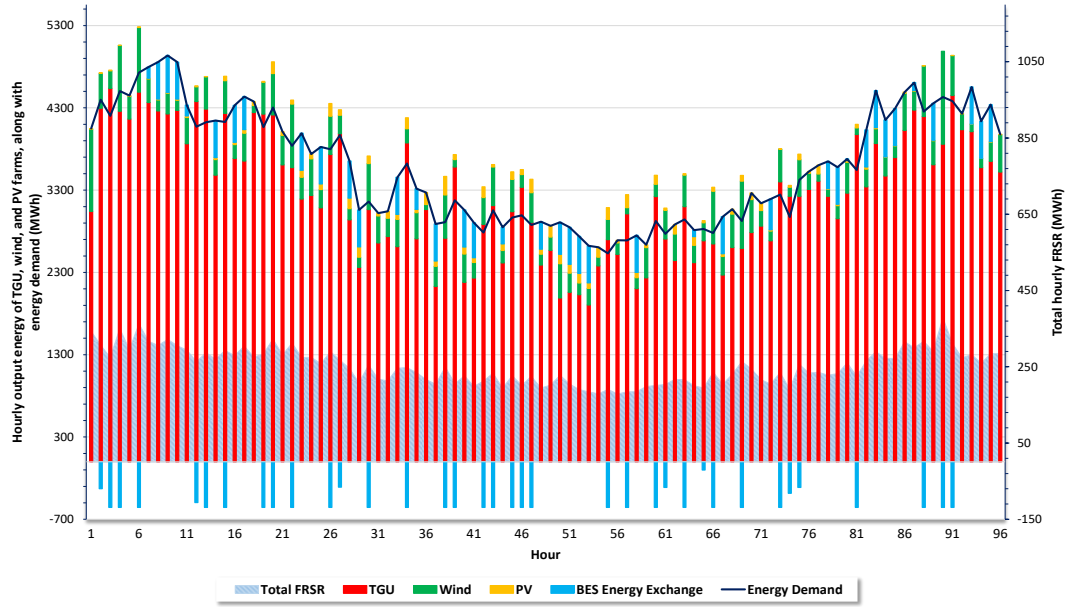

Fig. 10. The hourly share of all units in supplying energy demand at the last planning stage of Case 5

## VI. CONCLUSION

This paper presented a multi-stage model for Transmission, Generation, and battery energy Storage Expansion Planning (TGSEP) considering Renewable Portfolio Standard (RPS) and Low-Carbon Policy (LCP). To capture the short-term uncertainties of load demand and Renewable Energy Sources (RESs), a hierarchical clustering method is developed. Also, the upward and downward Flexible Ramp Spinning Reserve (FRSR) are incorporated in the proposed model. The major findings of this work are summarized as follows. 1) Joint expansion planning of transmission network and generation technologies facilitate the integration of RESs and Battery Energy Storage (BES) devices. 2) According to the obtained results, the incorporation of BES devices in the integrated expansion planning of transmission and generation defers the Thermal Generation Units (TGUs) investment, minimizes the total RESs curtailment, reduces the transmission congestion to facilitate the integration of RESs, and consequently reduces the total operation, emission, and planning cost. 3) In addition to the impact of LCP on decreasing the growth rate of $CO_2$ emission, the BES devices can noticeably reduce $CO_2$ emission. 4) Considering the sitting and sizing of RESs under RPS policy in the integrated expansion planning of transmission, generation, and BES devices reduces the share of TGUs in load demand supplying, and the total operation and emission costs. 5) Utilizing a proper hierarchical clustering method and modeling the upward and downward FRSR are essential to capture the uncertainties of load demand and RESs accurately. 6) The accelerated Benders Dual Decomposition (BDD) method is able to solve the proposed multi-stage TGSEP model with a tractable complexity, and facilitate the simultaneous optimization of both investment and operational variables by decomposing the original problem to a master problem and several dual sub-problems. The integration of DTR system and other energy storage devices, e.g., pumped storage hydro units, and their impacts on expansion planning is an open question that can be addressed in future researches.